\documentclass[aps,pra,reprint,superscriptaddress,showpacs]{revtex4-1}
\usepackage[utf8]{inputenc}
\usepackage{color}
\usepackage{graphicx}
\usepackage{enumitem}
\usepackage{amssymb,amsmath,amsthm,amsfonts}
\usepackage{mathtools,braket,array}
\usepackage{float}
\newcolumntype{C}[1]{>{\centering\let\newline\\\arraybackslash\hspace{0pt}}m{#1}}

\begin{document}
\title{Realizing two-qubit gates through mode engineering on a trapped-ion quantum computer}
\author{Ming Li}
\email{Present address: Atom Computing, Inc., Berkeley, CA 94710, USA}
\affiliation{IonQ, College Park, MD 20740, USA}
\author{Nhung H. Nguyen}
\affiliation{Joint Quantum Institute and Department of Physics, University of Maryland, College Park, MD 20742, USA}
\affiliation{Department of Physics, University of Maryland, College Park, MD 20742, USA}
\author{Alaina M. Green}
\affiliation{Joint Quantum Institute and Department of Physics, University of Maryland, College Park, MD 20742, USA}
\affiliation{Department of Physics, University of Maryland, College Park, MD 20742, USA}
\author{Jason Amini}
\affiliation{IonQ, College Park, MD 20740, USA}
\author{Norbert M. Linke}
\affiliation{Joint Quantum Institute and Department of Physics, University of Maryland, College Park, MD 20742, USA}
\affiliation{Department of Physics, University of Maryland, College Park, MD 20742, USA}
\affiliation{
Duke Quantum Center and Department of Physics, 
Duke University, Durham, North Carolina 27708, USA}
\author{Yunseong Nam}
\affiliation{IonQ, College Park, MD 20740, USA}
\affiliation{Department of Physics, University of Maryland, College Park, MD 20742, USA}
\date{\today}

\begin{abstract}
Two-qubit gates are a fundamental constituent of a quantum computer and typically its most challenging operation. In a trapped-ion quantum computer, this is typically implemented with laser beams which are modulated in amplitude, frequency, phase, or a combination of these. The required modulation becomes increasingly more complex as the quantum computer becomes larger, complicating the control hardware design. Here, we develop a simple method to essentially remove the pulse-modulation complexity by engineering the normal modes of the ion chain. We experimentally demonstrate the required mode engineering in a three ion chain. This opens up the possibility to trade off complexity between the design of the trapping fields and the optical control system, which will help scale the ion trap quantum computing platform.
\end{abstract}

\maketitle

\section{Introduction}

Quantum computers are limited in their computational power, mainly due to their immature underlying technologies. The cost of two-qubit gates typically dominates the error budget of a quantum circuit and represents a bottleneck for increasing the power of
quantum computational systems \cite{ballance2016,gaebler2016,levine2018,google2019}.
This true for trapped-ion quantum computers (TIQC), which are a leading platform for quantum computing. Two-qubit gates rely on the shared motional modes of the harmonically confined ion chain. Spin-motion couplings are driven with laser beams or microwave pulses, with the former being the preferred method for multi-qubit devices due to the easier addressability of individual ions \cite{nagerl1999}. As a result, a lot of the control complexity in a contemporary TIQC stems from the necessity to precisely control the frequency, amplitude, and phase of an array of laser beams. Reducing this complexity would greatly improve the robustness and scalability of the system and enable miniaturization.

In this paper, we take a first step towards addressing
this problem by developing a method to trade off the 
complexity in modulating the optical pulses against implementing more tailored static trapping potentials. We engineer the motional mode spectrum such that entangling operations can be driven with single-frequency pulses at constant amplitude and verify our method by demonstrating it on a small ion chain. 
Our successful proof-of-principle demonstration paves the way for scaling this gate scheme to larger ion chains at the cost of more complex mode engineering in a co-design approach with the trap electrode.

\section{Two-qubit gates on a trapped-ion quantum computer}

In this section, we briefly review the two-qubit XX gate,
typically used in a trapped-ion quantum computer.
Specifically, we assume a hardware configuration
of a linear chain of ions, each addressed with an individual laser beam.

An XX-gate between 
two trapped-ion qubits $i$ and $j$ is defined by the unitary operator
\begin{equation}
    {\rm XX}\left(\theta_{ij}\right) = e^{-i\theta_{ij}\left(
      \sigma_x^i\otimes\sigma_x^j\right)/2} \;, 
\label{eq:XX}
\end{equation}
where $\theta_{ij}$ denotes the degree of entanglement and
$\sigma_x^{i,j}$ are the Pauli-$x$ matrices acting on the qubit space $i$ and
$j$, respectively. To implement such a gate, the standard approach is the M{\o}lmer-S{\o}rensen (MS)
\cite{MS-1, MS-2} protocol, which uses transient coupling between the qubit and the shared motional Hilbert spaces.
Ideally, the two spaces are decoupled from each other at the end of the gate. The residual coupling, $\alpha$, between motional modes $p$ and ions $i$ and $j$ after the gate operation is given by  
\begin{align}
    \alpha :=& \frac{4}{5} \sum_{p=1}^N \coth\frac{\hbar\omega_p}{2k_BT_p}
    ({\eta_p^i}^2+{\eta_p^j}^2)\left|\int_0^\tau g(t)
    e^{i(\omega_p t + \phi_p)} {\rm d}t\right|^2,
\label{eq:alpha}
\end{align}
where $N$ is the number of modes involved in the coupling,
$g(t)$ is the shape of the pulse that illuminates individual ions,
$\eta_p^{i,j}$ are the Lamb-Dicke parameters,
$\tau$ is the total gate time,
$\omega_p$ is the angular mode frequency, 
$\phi_p$ is the initial phase associated with mode $p$ at time $t=0$, $T_p$ is the temperature of the $p$-th mode,
$\hbar$ is the reduced Planck constant, and
$k_B$ is the Boltzmann constant~\cite{GREENPAP}.
This is equivalent to the infidelity caused by residual motional entanglement and for an error budget of $\epsilon$ we require $\alpha < \epsilon$.

$\theta_{ij}$ is given by 
the sum of time-ordered double integrals, i.e.,
\begin{align}
    \chi_{ij} :=& \sum_{p=1}^{N} \eta_p^i\eta_p^j
    \int_0^\tau {\rm d}t_2 \int_0^{t_2} {\rm d}t_1
    g(t_2) g(t_1) \sin\left[\omega_p(t_2-t_1)\right] \nonumber\\
    =& \theta_{ij}/4 \;.
\label{eq:chi}
\end{align}
A fully entangling XX-gate requires $\theta_{ij} = \pi/2$, 
or $\chi_{ij} = \pi/8$.
An inexact implementation that induces uncertainty in $\chi_{ij}$ also 
leads to gate infidelity.

To meet these two constraints, Eqs. (\ref{eq:alpha}) and (\ref{eq:chi}), many different schemes have 
been developed.
For instance, given a mode-frequency spectrum, 
the pulse function $g(t)$ is varied through amplitude modulation (AM)~\cite{AM,transverse2006,roos2008,choi2014,zarantonello2019}, 
frequency modulation (FM)~\cite{FM}, 
phase modulation (PM)~\cite{PM,hayes2012,shapira2018,milne2020}, or combinations of these \cite{shapira2018} including power optimal AMFM~\cite{AMFM} schemes.
An important requirement to all these schemes is that 
one needs to ensure that the laser field seen by the ions follows the designed pulse shape faithfully.
As this pulse shape becomes increasingly complex, 
which is typically the case as we add more qubits for more powerful
quantum computation or decrease the gate duration for faster quantum
computation, this becomes challenging in practice.
Sampling rate, frequency and amplitude resolution, rise and fall time, and device signal distortion can limit the accuracy of the modulation.

\section{Motional mode engineering for a two-qubit gate pulse}

In this section, we explore the idea of engineering the
motional modes themselves to help simplify the pulse shape $g(t)$.
Note the pulse shaping schemes discussed in the previous section
assume the motional modes as a fixed input parameters.
Here we trade the pulse-shape complexity for 
overhead in engineering the mode spectrum.

\subsection{Requirements}
To satisfy (\ref{eq:alpha}) and (\ref{eq:chi}) in our mode-engineering based approach,
we require the mode frequency spectrum to obey the

 \begin{description}
[noitemsep,topsep=0pt]
 \item[Mode Engineering Condition (MEC)]
 {\noindent $\omega_p\tau/4 = k_p\pi$, where $k_p$ are
 positive integers.}
\label{condition1}
\end{description}
 
 This means the trap 
and electrode voltages must be designed such that the mode structure fulfills the \textbf{MEC} for an acceptable gate time $\tau$.
Once the \textbf{MEC} is fulfilled, the pulse shapes required to realize a gate simplify dramatically. For example, we can choose the pulse $g(t)$ to obey
\newline
\begin{description}
 \item[Prime 1] $g(t+\tau/2) = -g(t)$ for $t \in [0, \tau/2]$.
\label{condition2}
\end{description}
Such a pulse corresponds to two consecutive sections of the same arbitrary shape, such as a square pulse, but with opposite phases. 
Using \textbf{MEC} and \textbf{Prime 1}, while allowing for an error
$|\delta k_p| \ll 1$ in mode frequency 
$\omega_p = 4(k_p + \delta k_p)\pi/\tau$, 
we can rewrite the integral in (\ref{eq:alpha}) as
\begin{widetext}
\begin{align}
    \int_0^\tau g(t)e^{i(\omega_p t + \phi_p)} {\rm d}t = &
     \int_0^{\tau/2} g(t)e^{i(\omega_p t + \phi_p)} {\rm d}t 
     + \int_{\tau/2}^\tau g(t)e^{i(\omega_p t + \phi_p)} {\rm d}t \nonumber\\
    = &
     \int_0^{\tau/2} g(t)e^{i(\omega_p t + \phi_p)} {\rm d}t
     + \int_{\tau/2}^\tau g(t)e^{i[\omega_p (t-\tau/2) 
     + \omega_p\tau/2 + \phi_p]} {\rm d}t \nonumber\\
    = &
     \int_0^{\tau/2} g(t)e^{i(\omega_p t + \phi_p)} {\rm d}t
     - \int_0^{\tau/2} g(t)e^{i(\omega_p t + \phi_p 
     + 2\delta k_p\pi)} {\rm d}t \nonumber\\
    = & (1-e^{2i\delta k_p\pi}) 
     \int_0^{\tau/2} g(t)e^{i(\omega_p t + \phi_p)} {\rm d}t \;.
\label{eq:integral}
\end{align}
\end{widetext}
As $\delta k_p \to 0$, Eq.~(\ref{eq:integral}) vanishes, 
upper-bounded by $\mathcal{O}[\max(|\delta k_p|)]$.
As long as $|\chi_{ij}|>0$ in (\ref{eq:chi}), we can scale $g(t)$ to satisfy the
gate requirements (\ref{eq:alpha}) and (\ref{eq:chi}) exactly for $\delta k_p= 0$. 
Therefore, \textbf{MEC} and \textbf{Prime 1} are sufficient to implement an XX gate in principle.

\subsection{Pulse simplification and power optimization}
To further streamline the pulse while also lowering the power requirement,
we propose the following simple procedure.
We choose the pulse to further satisfy
\begin{description}
 \item[Prime 2] $g(t) = \Omega\sin\left(\frac{2l\pi t}{\tau}\right)$
     for $t \in [0, \tau/2]$ where $l$ is a positive integer and $\Omega$ is the Rabi frequency.
\end{description}
This simplifies the generation of the optical signal,
as the pulse has constant amplitude, is continuous throughout, and is monotone with the frequency set by $2\pi l/\tau$.
Combining \textbf{Prime 1} and \textbf{Prime 2}, we notice that  we can separate these pulses into groups with odd and even $l$.
We can derive an analytic expression for $\chi_{ij}$ given by
\begin{widetext}
\begin{align}
    \chi_{ij} = 
    \begin{dcases*}
     \frac{\Omega^2\tau^2}{2\pi}
      \sum_{p=1}^{N} \eta_p^i\eta_p^j \frac{k_p}{4k_p^2 - l^2} \;,
       & \text{if $\nexists \; p=1,\cdots,N$ such that $k_p = l/2$,} \\
     \frac{\Omega^2\tau^2}{2\pi} \left(
      \eta_{p_l}^i\eta_{p_l}^j\frac{3}{8 l} + 
      \sum_{\substack{p=1\\p\ne p_l}}^{N} 
       \eta_p^i\eta_p^j \frac{k_p}{4k_p^2 - l^2}
      \right)  \;,
       & \text{if $\exists \; p=1,\cdots,N$ such that $k_p = l/2$.}
    \end{dcases*}
\label{eq:chi_analytic}
\end{align}
\end{widetext}
Here, $p_l$ is the index of a specific mode that satisfies $2k_{p_l} = l$.
For a given set of Lamb-Dicke parameters and mode
frequencies, it is straightforward to pick an $l$ satisfying Eq. (\ref{eq:chi})  with minimal $\Omega$, which corresponds to the lowest gate power, by maximizing the quantity $|\chi_{ij}/\Omega^2\tau^2|$ according to Eq. (\ref{eq:chi_analytic}). In practice, $l$ should be chosen as a compromise between gate power and robustness to error, which we discuss next.

\subsection{Residual motional coupling}

We calculate the largest $|\delta k_p|$, for a given error budget $\varepsilon$.
To start, assuming non-zero $\delta k_p$, we use \textbf{Prime 2} to write
\begin{widetext}
\begin{align}
    \alpha = 
    \begin{dcases*}
     \frac{l^2\Omega^2\tau^2}{5\pi^2}
     \sum_{p=1}^N \left[\coth\left(\frac{\hbar\omega_p}{2k_BT_p}\right)\right]
     ({\eta_p^i}^2+{\eta_p^j}^2)\left|
      \frac{e^{i\phi_p}(e^{4i\delta k_p\pi}-1)}{4(k_p+\delta k_p)^2 - l^2}
     \right|^2 \;,
     & \text{if $l$ is odd,} \\
    \frac{l^2\Omega^2\tau^2}{5\pi^2}
     \sum_{p=1}^N \left[\coth\left(\frac{\hbar\omega_p}{2k_BT_p}\right)\right]
     ({\eta_p^i}^2+{\eta_p^j}^2)\left|
      \frac{e^{i\phi_p}(e^{2i\delta k_p\pi}-1)^2}{4(k_p+\delta k_p)^2 - l^2}
     \right|^2 \;,
     & \text{if $l$ is even.}
    \end{dcases*}
\label{eq:alpha_precool}
\end{align}
\end{widetext}
Assuming the motional modes are sufficiently cooled to average phonon number $\bar{n}<1$,
which is readily achieved by sideband cooling~\cite{SimCool}, we may let
$\coth\left(\frac{\hbar\omega_p}{2k_BT_p}\right) < 2$. Dropping now the modulus-one term
$e^{i\phi_p}$, we can bound $\alpha$ by
\begin{widetext}
\begin{align}
    \alpha < 
    \begin{dcases*}
     \frac{2l^2\Omega^2\tau^2}{5\pi^2}
     \sum_{p=1}^N
     {(\eta_p^i}^2+{\eta_p^j}^2)\left|
      \frac{e^{4i\delta k_p\pi}-1}{4(k_p+\delta k_p)^2 - l^2}
     \right|^2 \;,
     & \text{if $l$ is odd,} \\
    \frac{2l^2\Omega^2\tau^2}{5\pi^2}
     \sum_{p=1}^N
     ({\eta_p^i}^2+{\eta_p^j}^2)\left|
      \frac{(e^{2i\delta k_p\pi}-1)^2}{4(k_p+\delta k_p)^2 - l^2}
     \right|^2 \;,
     & \text{if $l$ is even.}
    \end{dcases*}
\end{align}

For small $\delta k_p$ we can expand the right-hand
side,

\begin{align}
    \alpha < 
    \begin{dcases*}
     \frac{32l^2\Omega^2\tau^2}{5}
     \sum_{p=1}^N
     ({\eta_p^i}^2+{\eta_p^j}^2)
     \frac{\delta k_p^2}{(4k_p^2 - l^2)^2}
     + \mathcal{O}(\delta k_p^3) \;,
     & \text{if $l$ is odd,} \\
    \alpha_0 + \frac{32l^2\Omega^2\tau^2\pi^2}{5}
     \sum_{\substack{p=1\\p\ne p_l}}^N
     ({\eta_p^i}^2+{\eta_p^j}^2)
     \frac{\delta k_p^4}{(4k_p^2 - l^2)^2}
     + \mathcal{O}(\delta k_p^5)\;,
     & \text{if $l$ is even,}
    \end{dcases*}
    \label{eq:alpha_analytic}
\end{align}
where
\begin{align}
    \alpha_0 = 
    \begin{dcases*}
     0 \;,
       & \text{if $\nexists \; p=1,\cdots,N$ that $k_p = l/2$,} \\
     \frac{2\Omega^2\tau^2\pi^2}{5}({\eta_{p_l}^i}^2+{\eta_{p_l}^j}^2)
     \delta k_{p_l}^2 + \mathcal{O}(\delta k_p^3) \;,
       & \text{if $\exists \; p=1,\cdots,N$ that $k_p = l/2$.}
    \end{dcases*}
    \label{eq:alpha0}
\end{align}
\end{widetext}
The gate is robust to deviation from the \textbf{MEC}, resulting either from imperfect mode engineering or mode drift, because the leading order in $\delta k_p$ is quartic at best, for even $l$ if there is no $p$ such that $k_p = l/2$, and quadratic at worst, otherwise.

\begin{figure}[t!]
    \centering
    \includegraphics[trim=10 0 0 0, clip, width=1.05\columnwidth]{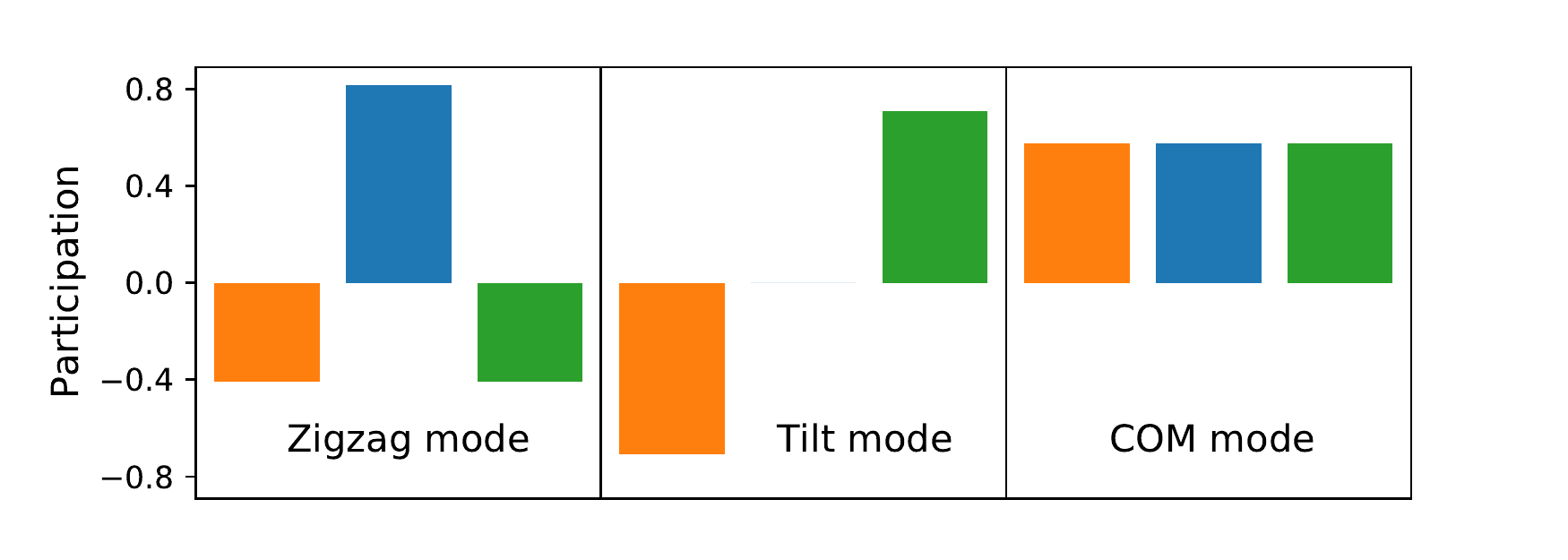}
    \caption{Example mode participation for a three-ion chain.
    Color orange represents ion 1, blue represents ion 2,
    and green represents ion 3. The zig-zag mode has the
    lowest mode frequency while the center-of-mass (COM) mode has the
    highest.}
    \label{fig:part}
\end{figure}

\begin{figure}[h]
    \centering
    \includegraphics[width=1.05\columnwidth]{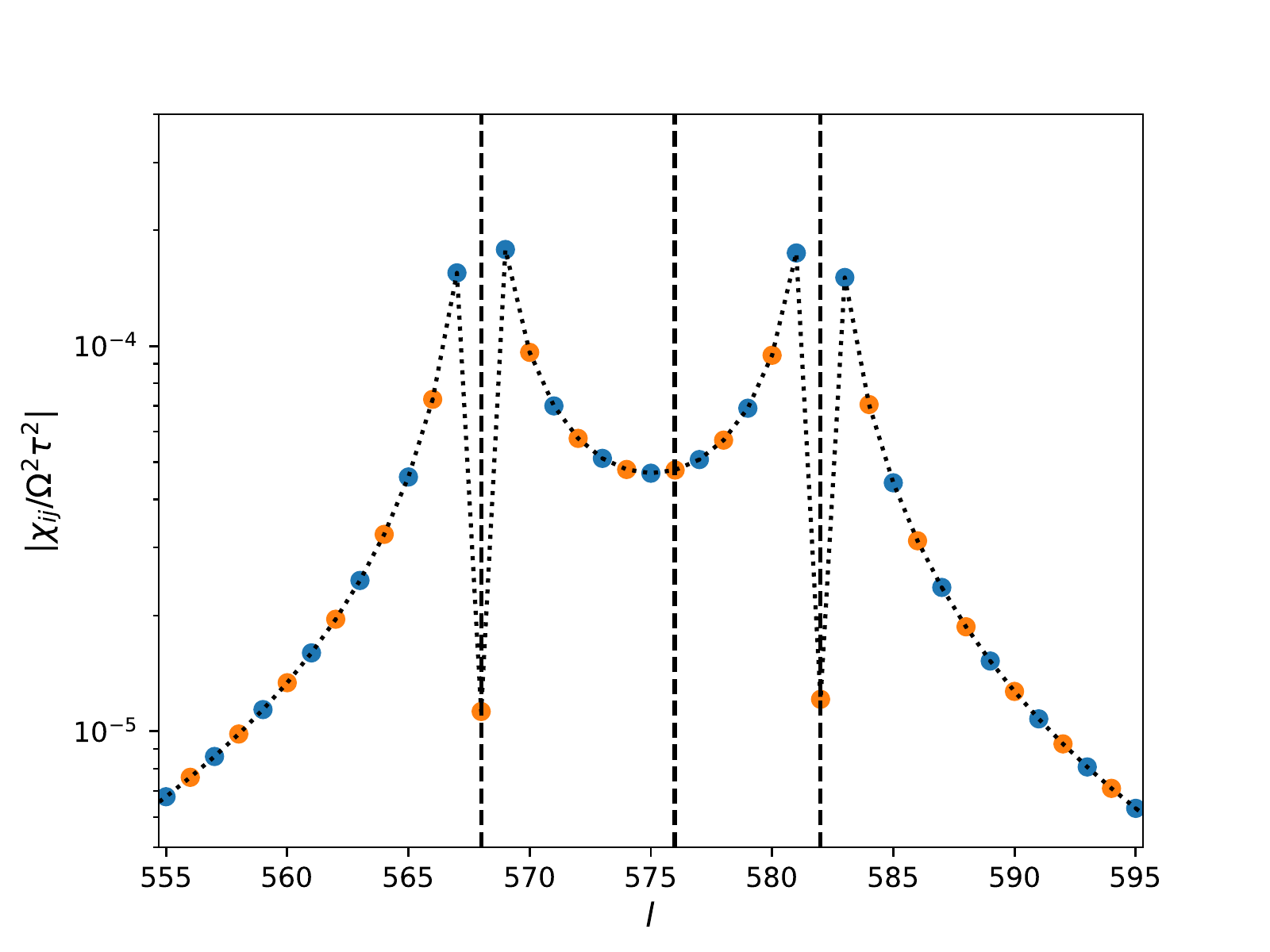} 
    \caption{$|\chi_{ij}/\Omega^2\tau^2|$ versus $l$
    in \textbf{Condition 4} for an XX-gate implementation
    between ions (1,2) using the ideal mode participation. The values for even $l$s are shown
    as filled orange circles and the ones for odd $l$s are
    shown as filled blue circles. The vertical black dashed-lines
    correspond to the $l = 2k_p$ of the ideal mode frequencies.
    We pick $l = 569$ as the odd $l$ pulse and $l = 570$
    as the even $l$ pulse for their lowest power requirement
    respectively.}
    \label{fig:mat}
\end{figure}

For a given error bound $\varepsilon$ on $\alpha$, 
we see that the requirement for $\delta k_p$ is
much less stringent for pulses with even $l$ than pulses with odd $l$, if there is no $p$ such that $k_p = l/2$.
However choosing such an even $l$ typically requires larger $\Omega$ and hence higher power. Therefore, in practice, the choice will depend on the specific experimental error budget. In the following we consider both even and odd $l$ for completeness.

\section{Experimental implementation}

\begin{figure}[h]
    \centering
    \includegraphics[width=1.05\columnwidth]{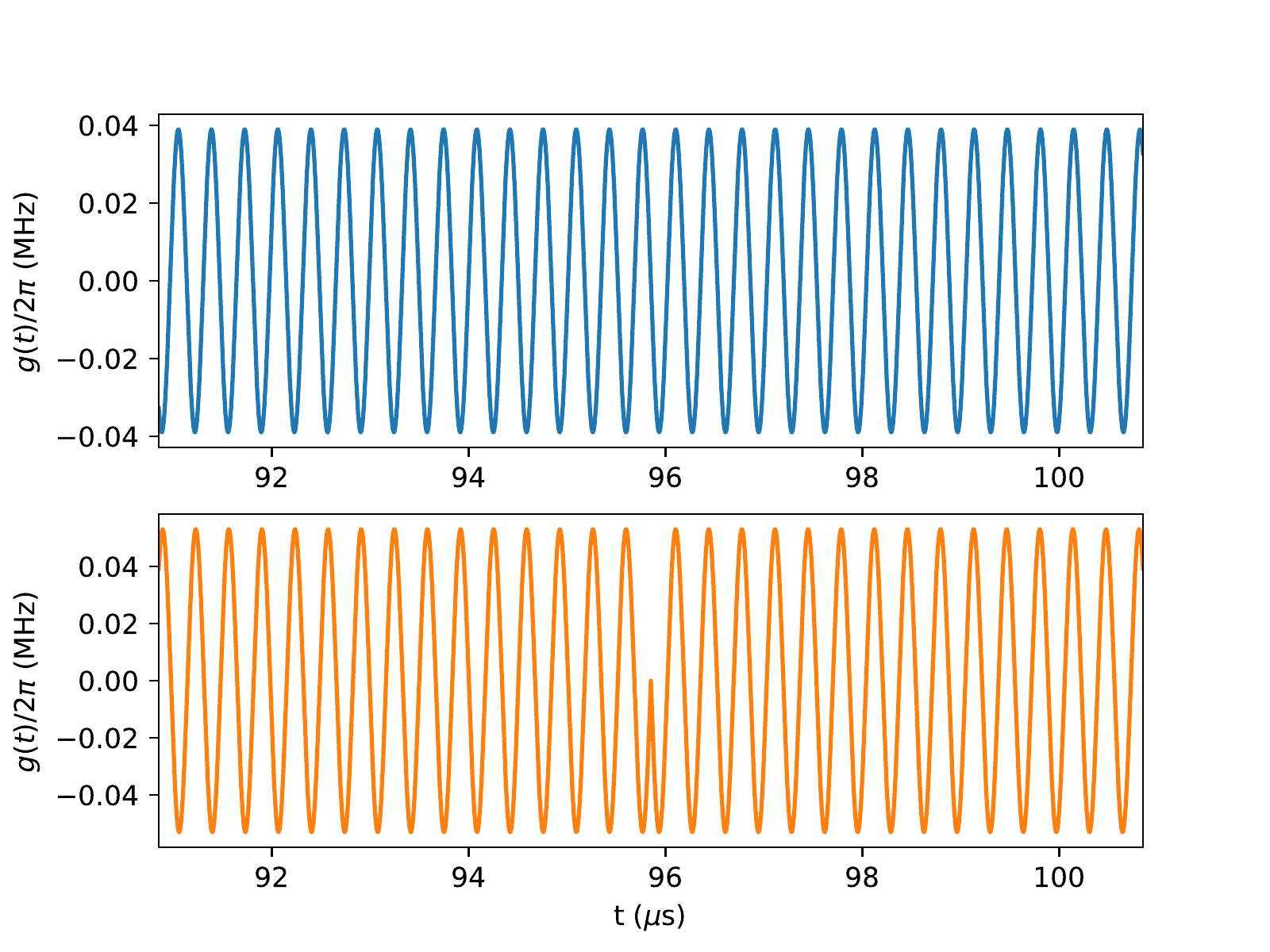} 
    \caption{Pulses of odd (upper), $l = 569$, and even (lower), $l = 570$, parities that implement an XX-gate between ions (1,2). The pulses were synthesized using the ideal mode participation.
    The pulse with odd $l$ is infinitely
    differentiable at the half way point of the gate at
    $t = 95.85 \mu$s while the pulse with even $l$ has a 
    cusp.}
    \label{fig:pulses}
    \vspace{-0.1in}
\end{figure}
The experiment is performed with three $^{171}\rm{Yb}^{+}$ ions held in a linear Paul trap.
We define the qubit state $\ket0$ to be $\ket{F=0,m_F=0}$ and $\ket1$ to be $\ket{F=1,m_F=0}$ in the $^{2}S_{1/2}$ ground level manifold. The ions are Doppler-cooled before being further cooled to the motional ground state via resolved sideband cooling. The qubits are then initialized to $\ket0$ via optical pumping. The ions' internal states and collective motions are manipulated with a pair of counter-propagating Raman beams, one of which is split into three beams for individual addressing \cite{debnath2016}. At the end of the experiment, the internal state is read out by state-dependent fluorescence \cite{olmschenk2007} collected via a multi-channel photo-multiplier tube.

\begin{figure*}[t!]
\centering
\hspace{-0.1in}
\includegraphics[width=\textwidth]{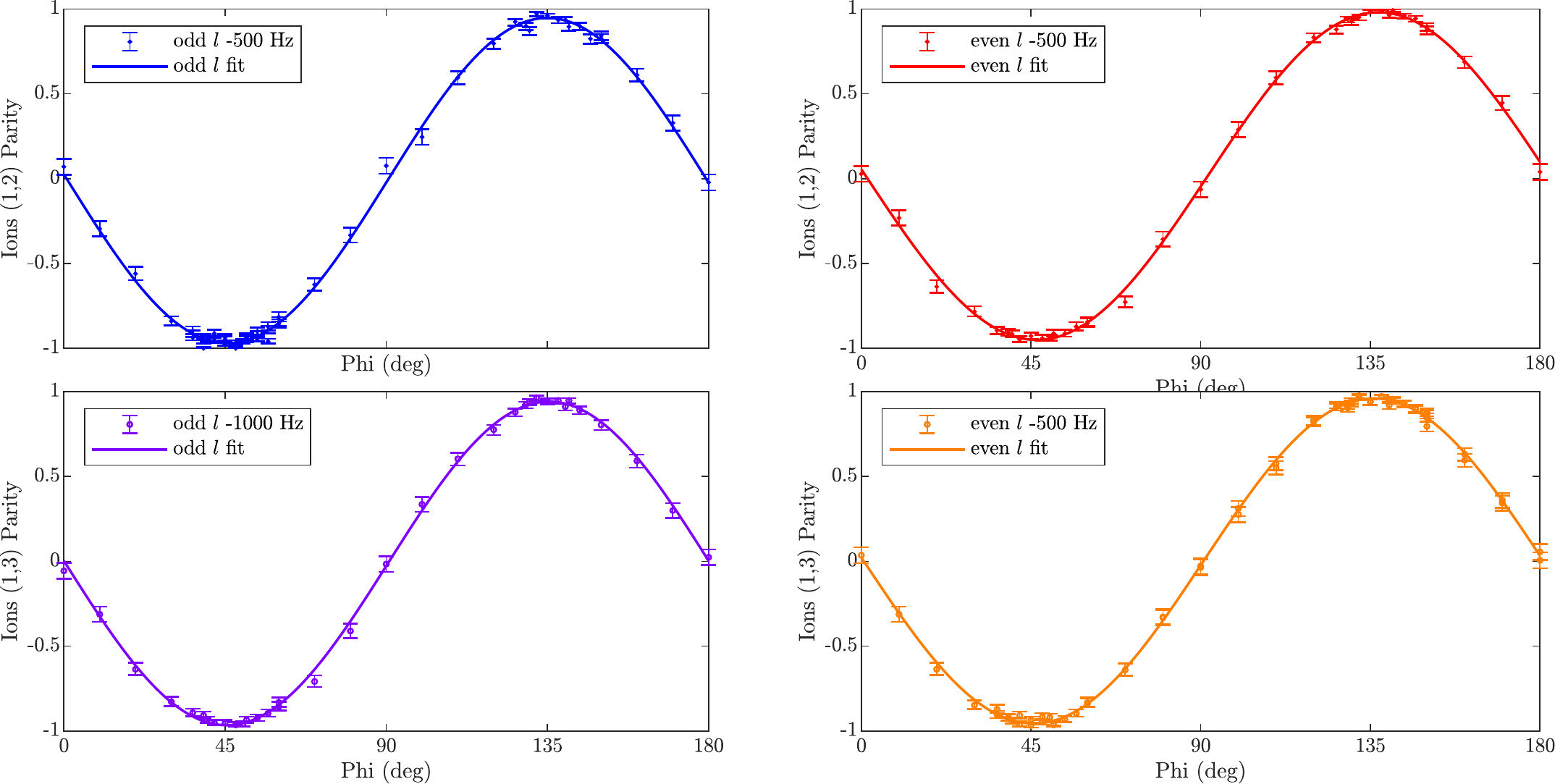}
\caption{Contrast of the parity for pulses on pairs (1,2) and (1,3) with odd $l$ (without phase flip) and even $l$ (with phase flip).} 
\label{fig:exp-high-fid}
\end{figure*}

We implement XX gates between ion pairs (1,2) and (1,3).
The experimentally measured radial mode frequencies  are $2\pi\times2.963$ MHz (zig-zag mode), $2\pi\times3.005$ MHz (tilt mode), and $2\pi\times3.036$ MHz (center-of-mass mode). Day-to-day drifts are on the order of a few kHz, while hour-to-hour drifts are on the order of a few hundred Hz. To approximately satisfy the \textbf{MEC} while minimizing the gate time, we choose $\tau=191.700$~$\mu$s, which makes $k_p = 284$, $288$, and $291$, respectively.
We use the ideal mode participation, illustrated in Fig.~\ref{fig:part},  for the Lamb-Dicke parameters and apply \textbf{Primes 1} and \textbf{2} to obtain
$g(t)$. This allows us to use (\ref{eq:chi_analytic})
to compute $|\chi_{ij}/\Omega^2\tau^2|$ for different
choices of $l$, shown in Fig.~\ref{fig:mat} for the ion pair (1,2). We pick the maxima, $l = 570$ and $569$ for even and odd $l$, respectively, which corrrespond to optimal gate power. Following the same steps, we pick $l=578$ and $577$ for ion pair (1,3).
Figure~\ref{fig:pulses} shows the middle section of the pulses for pair (1,2), featuring a phase-flip for $l = 570$ at the half-way point of the pulse,
which does not exist for $l = 569$. The pulses successfully drive high-fidelity entangling gates for both pairs. The fidelity is determined by the average even state population and the parity contrast, all of which are shown in Table~\ref{tab:fid}. The even state population is the sum of population in $\ket{00}$ and $\ket{11}$ of the maximally entangled state, which is the result of applying an $\rm{XX}(\pi/4)$ on $\ket{00}$. The parity contrast, as shown in  Fig.~\ref{fig:exp-high-fid}, are obtained from measuring the entangled state in different basis by applying 
a $pi/2$ rotation with a phase $phi$
on both ions followed by a projective measurement on the $Z$-basis.

\begin{table}[h]
\caption{Parity contrast, even state population and fidelity for pairs (1,2) and (1,3) at optimal added detuning. The parity contrast for these pulses are shown in Fig.~\ref{fig:exp-high-fid}.}
\begin{tabular}{l c C{2.0cm} c C{1.5cm} c}
\hline
Pairs & $l$  & Added detuning (Hz) & Contrast & Even population & Fidelity \\
\hline
(1,2) & 569  & -500      & 0.958(9)  & 0.978(3) & 0.968(9)  \\
(1,2) & 570 & -500      & 0.966(8)  & 0.974(3) & 0.970(9)  \\
(1,3) & 577  & -1000      & 0.954(9) & 0.983(2) & 0.97(1) \\
(1,3) & 578 & -500      & 0.958(9) & 0.986(3) & 0.97(1)  \\
\hline
\end{tabular}
\label{tab:fid}
\end{table}

\begin{figure}[t!]
\centering
\hspace{-0.1in}
\includegraphics[width=0.45\textwidth]{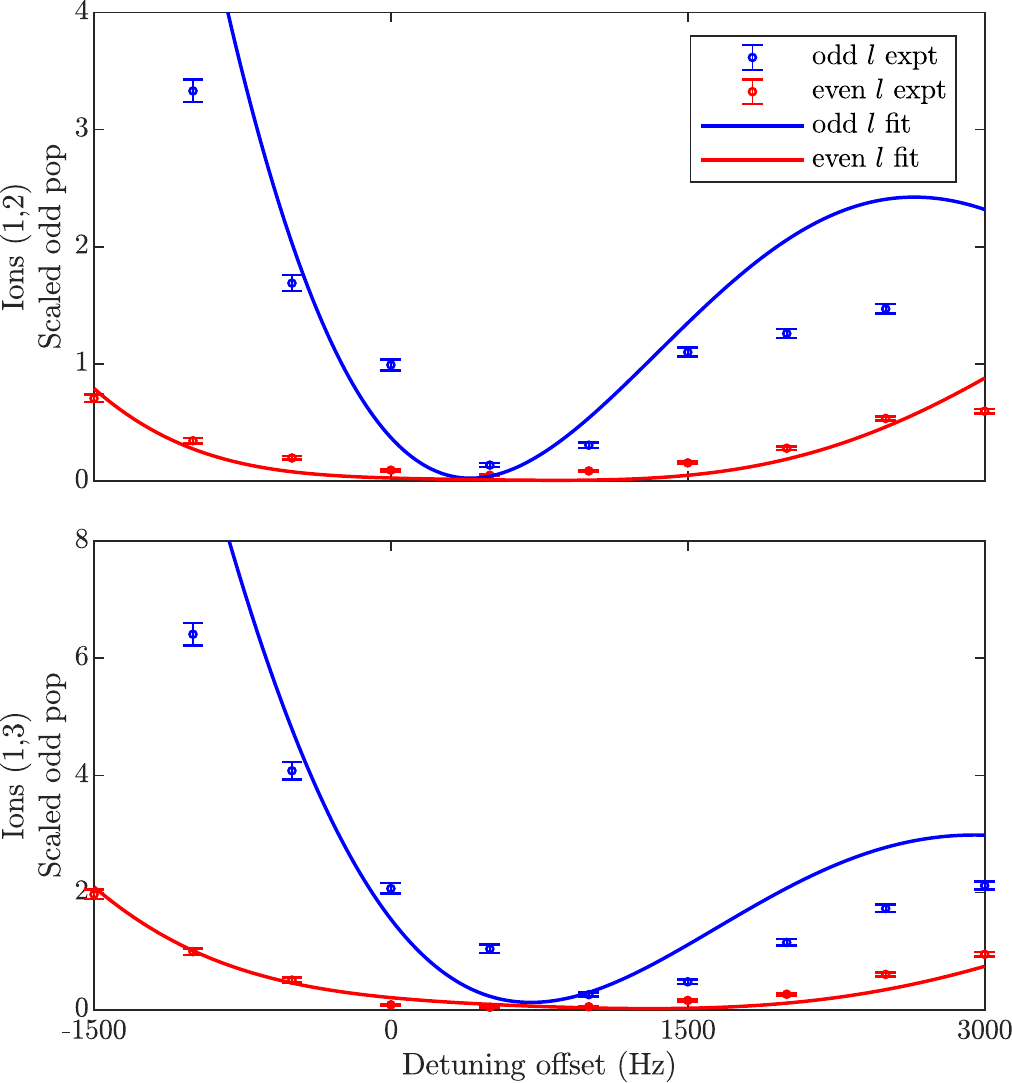}
\caption{Robustness of the phase space closure against frequency drift for pulses with even $l$ (with phase flip) and odd $l$ (without phase flip) as shown by the scaled odd state population for for the ion pairs (1,2) and (1,3). The scaled odd state population is the measured odd state population scaled by the square of Rabi frequency $\Omega^2$ for a maximally entangled gate where $\theta_{ij}=\frac\pi2$. The lines are fits to the theoretical model.} 
\label{fig:exp}
\end{figure}

We next examine the stability of the gate with respect to errors in mode frequency engineering and mode drift by measuring the odd population for different 
gate frequency offsets $\delta\omega$. At $\delta\omega = 0$ the infidelity due to residual
motional couplings, as quantified by $\alpha$, is caused by the static error of mode engineering only.
For $\delta\omega\neq0$ it approximates the errors caused by common-mode deviation of the experimental mode frequency values. 
For the experiments, we prepare $\ket{00}$ as our initial state, apply an XX$(\pi/4)$ gate using the synthesized pulses, then measure the odd state population, i.e. $\ket{01}$ and $\ket{10}$, which is a measure of phase space closure.
We observe how the odd state population changes with $\delta\omega$. Since changing the gate detuning can change the value of $\chi_{ij}$, we calibrate the gate for every offset value to ensure an equal population for $\ket{00}$ and $\ket{11}$, and record the Rabi frequency needed. To compare the results across the different $\delta\omega$ values, we scale the odd state population by the square of the Rabi frequency needed for a maximally entangling gate. Figure~\ref{fig:exp} shows the results.
The blue and red data points represent pulses with odd $l$ and even $l$, respectively. The experimental parameters are given in Appendix \ref{app:stability}. 

To compare with theory, we numerically evaluate Eq.~(\ref{eq:alpha}) using the experimentally implemented pulses and perform a collective fit of all data points by only varying the three mode frequencies to minimize the root-mean-square of the differences between experimental data and theoretical predictions. The model uses experimentally measured $\eta_p^i$ and takes $\coth{(\hbar\omega_p/2k_BT_p)} = 1$, i.e. the average phonon occupation is assumed to be much smaller than one.
We attribute the remaining mismatch to slow drift of the mode frequencies over several hours, which is not accounted for in the collective fit.
As expected from Eqs. (\ref{eq:alpha_analytic}) and (\ref{eq:alpha0}), the pulses with even $l$ are much
more robust since the residual coupling error scales as $\delta\omega^4$, compared to the pulses with odd $l$, where it scales as $\delta\omega^2$. 
The additional requirement of $l \neq 2k_p$ for any mode $p$ is fulfilled in our examples, and does not restrict the solution space in general since it only requires the gate frequency be off-resonant with any of the modes.

The trade-off is a higher power requirement for even $l$, compared to odd $l$ solutions.
The pulses with even $l$ require $\Omega/2\pi = 0.0849$ MHz and $0.0592$ for $l=570$ and $578$, respectively, compared with $\Omega/2\pi = 0.0585$ MHz and $0.0454$ for $l = 569$ and $577$, respectively, an increase of about $30\%$.
Considering the additional power requirement may be prohibitive in 
terms of implementation, the choice between even vs. odd $l$ 
should be made carefully on a case by case basis, 
while fully taking the hardware limit into consideration.

Note that our experiments are performed on a blade trap where each DC blade has three segments.
Due to limited voltage control available in this trap, 
it is difficult to generate a scalable, analytical electrode-voltage solution
that will result in a linear ion chain that satisfies \textbf{MEC} 
for arbitrarily many ions. 
For the three-ion case, however, we can empirically optimize. We choose the RF frequency, RF and DC voltages so that the ions form a linear equally spaced crystal and micromotion is minimized.

\section{Outlook}

While our proof-of-principle demonstration of an XX-gate via mode engineering shows the viability of the technique, it was only realized in a three-ion chain and a trap that was not made for this purpose. For scaling up to systems involving more ions and hence more motional modes, we need to explore options for including the control elements with an eye toward the technical requirements
and challenges.

An ion trap sets the radial and axial mode structures of a chain of trapped ions through  a combination of the ion spacing and local potentials around each ion.  When working with the trapping potential, it is useful to expand the potential into real spherical harmonics $Y_{lm}(\theta,\phi)$ where $\theta$ is the polar angle, $\phi$ is the azimuthal angle, the polar axis is taken along the chain axis, and the origin is at the axial symmetry point of the chain. When combined with pseudopotential for a linear trap, $Y_{20}$ provides the initial harmonic confinement and $Y_{22}$ provides a net splitting between the two radial mode sets. Higher order terms become controls to manipulate the ion axial positions and the axial dependence of the radial confinement. Selection of $Y_{lm}$ with particular symmetries can greatly reduce the number of controls, for example, limiting the set to only those that generate axially symmetric electric fields and do not mix the two radial modes. While not a trivial optimization, it is one that can be approached iteratively to produce a particular mode structure. 

Creating complex potentials from this optimization above may require designing the trap explicitly with these potentials in mind. Multi-electrode surface traps ~\cite{NIST, HOA, phoenix2020, pino2021} provide arrays of electrodes for control of the potentials around the ions. With appropriate technological advancements, 
an increased number of trap electrodes and their shapes
may be leveraged to induce the desired potentials.
The design-space for tailoring a system optimized for mode engineering includes trade-offs between ion placement, micromotion, spatial twisting of mode participation, and mode spacing, which must be carefully considered. 

Furthermore, as quantum computing hardware scales to larger numbers of qubits, the traps will have to be designed as specialist devices even at the loss of some operational flexibility. There will then be tradeoffs in the designs and optimizing mode structure will need to be folded into these tradeoffs.  If a design choice in the trap can simplify the system operation through simplified gate operations, that may be a desirable tradeoff. This proposal should motivate new trap designs with electrodes shaped for producing designer potentials for the purpose of controlling the gate modes. 

Further adjustments to the ion potentials could be made
using potentials induced from non-trap sources as well. For example,
optical dipole potentials can further be used to modify potentials around each ion individually by employing optical tweezers arrays parallel to a surface trap~\cite{tweezers, IBK}.

\section*{Acknowledgement}
This material was based on work supported
by IonQ Inc., while M.L., and Y.N. were working
at IonQ Inc. Any opinion, finding, and conclusions or
recommendations expressed in this material are those of
the authors and do not necessarily reflect the views of
IonQ Inc.
This material is based upon work supported by the U.S. Department of Energy, Office of Science, National Quantum Information Science Research Centers, Quantum Systems Accelerator. N.M.L. received support from the National Science Foundation Quantum Leap Challenge Institutes (QLCI) (Grant No. OMA-2120757) and Physics Frontiers Centers (PFC) (Grant No. PHY-1430094). A.M.G. is supported by a Joint Quantum Institute Postdoctoral Fellowship.

\vspace{1cm}

\appendix

\section{Gates stability against mode drift
\label{app:stability}}
\noindent

For the stability data, the experimentally measured radial mode frequencies  are $2\pi\times2.964$ MHz (zig-zag mode), $2\pi\times3.006$ MHz (tilt mode), and $2\pi\times3.037$ MHz (center-of-mass mode). Day-to-day drifts are on the order of few kHz. To approximately satisfy the \textbf{MEC} while minimizing the gate time, we choose $\tau=192.946$~$\mu s$, which makes $k_p = 286$, $290$, and $293$ respectively.
We use the ideal mode participation, illustrated in Fig.~\ref{fig:part},  for the Lamb-Dicke parameters and apply \textbf{Prime 1} and \textbf{2} to
$g(t)$. This allows us to use (\ref{eq:chi_analytic})
to compute $|\chi_{ij}/\Omega^2\tau^2|$ for different
choices of $l$, shown in Fig.~\ref{fig:mat} for the ion pair (1,2). We pick the maxima, $l = 574$ and $573$ for even and odd $l$, respectively, which corrrespond to optimal gate power. Following the same steps, we pick $l=582$ and $581$ for ion pair (1,3).
Figure~\ref{fig:pulses} shows the middle section of the pulses for pair (1,2), featuring a phase-flip for $l = 574$ at the half-way point of the pulse,
which does not exist for $l = 573$. 

\begin{figure}[t!]
    \centering
    \includegraphics[width=1.05\columnwidth]{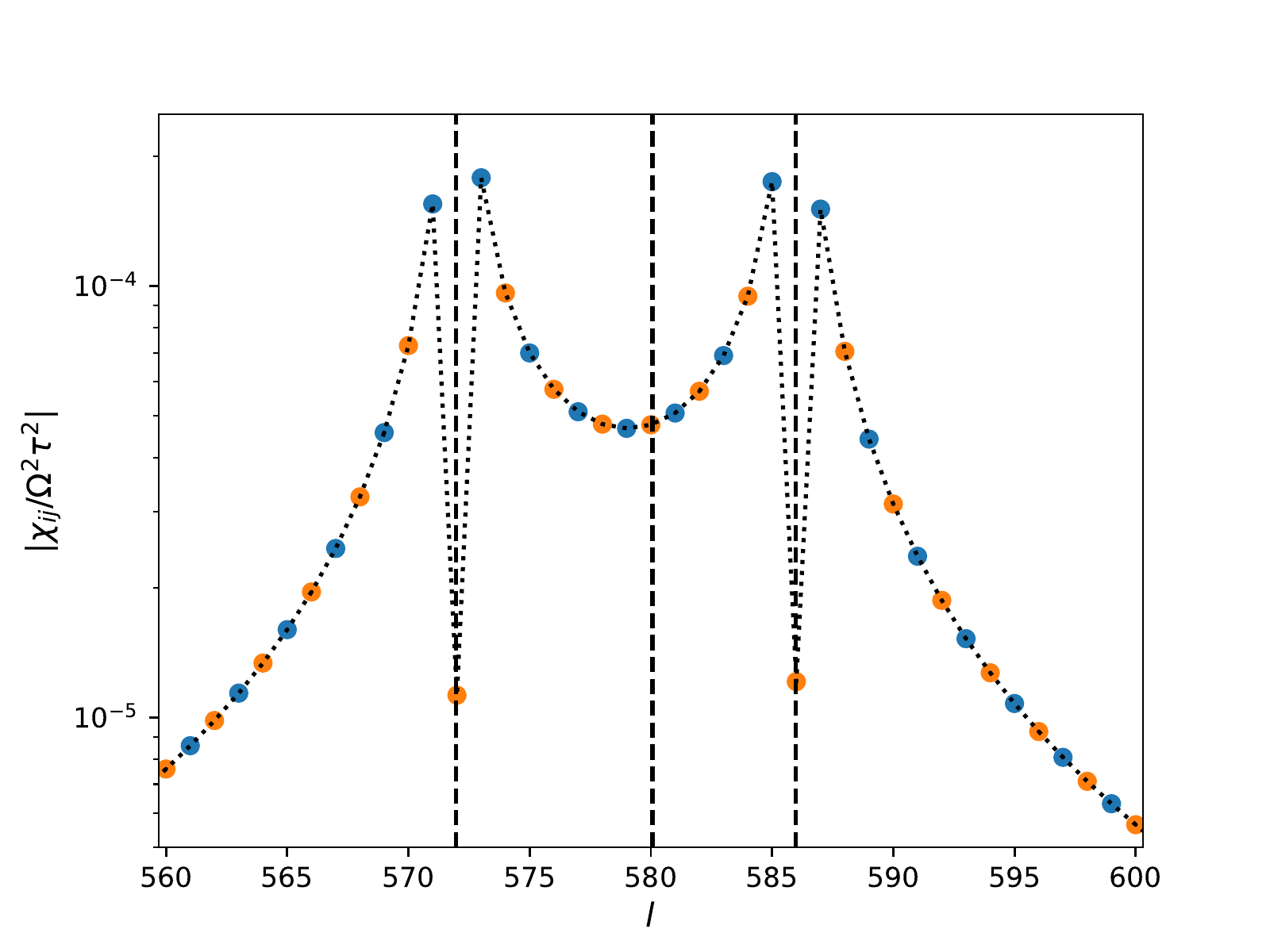} 
    \caption{$|\chi_{ij}/\Omega^2\tau^2|$ versus $l$
    in \textbf{Condition 4} for an XX-gate implementation
    between ions (1,2) using the ideal mode participation. The values for even $l$s are shown
    as filled orange circles and the ones for odd $l$s are
    shown as filled blue circles. The vertical black dashed-lines
    correspond to the $l = 2k_p$ of the ideal mode frequencies.
    We pick $l = 573$ as the odd $l$ pulse and $l = 574$
    as the even $l$ pulse for their lowest power requirement
    respectively.}
    \label{fig:mat}
\end{figure}

\end{document}